\documentclass{jkas}

\def\year{2018} 
\def\month{August} 
\def\volume{51} 
\def\issue{4} 
\def\beginpage{111} 
\def\endpage{117} 
\def\received{June 8, 2018} 
\def\accepted{August 1, 2018} 
\date{Received \received; accepted \accepted}

\usepackage{flushend} 
\usepackage[ps2pdf, breaklinks, colorlinks, citecolor=blue, linkcolor=blue, menucolor=blue, urlcolor=blue]{hyperref}

\title{
A Numerical Method to Analyze Geometric Factors of a Space Particle Detector Relative to Omnidirectional Proton and Electron Fluxes
}

\author[]{Sungmin~Pak}
\author[]{Yuchul~Shin}
\author[]{Ju~Woo}
\author[]{Jongho~Seon}

\affil[]{School of Space Research, Kyung Hee University, Yongin, Gyeonggi 17104, Korea; \email{jhseon@khu.ac.kr}}

\def\corrauthor{
J. Seon
}

\def\runningauthor{
Pak et al.
}

\def\runningtitle{
Geometric Factors of a Space Particle Detector relative to Omnidirectional Fluxes
}

\def\keywords{
radiation mechanisms: general --- instrumentation: detectors --- methods: data analysis --- methods: numerical
}

\def\abstracttext{
A numerical method is proposed to calculate the response of detectors measuring particle energies from incident isotropic fluxes of electrons and positive ions. The isotropic flux is generated by injecting particles moving radially inward on a hypothetical, spherical surface encompassing the detectors. A geometric projection of the field-of-view from the detectors onto the spherical surface allows for the identification of initial positions and momenta corresponding to the clear field-of-view of the detectors. The contamination of detector responses by particles penetrating through, or scattering off, the structure is also similarly identified by tracing the initial positions and momenta of the detected particles. The relative contribution from the contaminating particles is calculated using {\tt GEANT4} to obtain the geometric factor of the instrument as a function of the energy. This calculation clearly shows that the geometric factor is a strong function of incident particle energies. The current investigation provides a simple and decisive method to analyze the instrument geometric factor, which is a complicated function of contributions from the anticipated field-of-view particles, together with penetrating or scattered particles.
}

\begin{document}
\jkashead 


\section{Introduction\label{sec:intro}}

The response of detectors generally can be expressed in terms of the geometric factor. The definition of the geometric factor is given by (\citealt{sullivan1971})
\begin{eqnarray}
\label{eq:counting}
C = \frac{1}{T}\int_{t_0}^{t_0+T}dt\int_{S}d\vec\sigma\cdot\hat{r}\int_{\Omega}d\omega\int_{0}^{\infty}dE \nonumber\\ \times \sum_{\alpha}\varepsilon_{\alpha}(E,\vec\sigma,\omega,t)J_{\alpha}(E,\omega,\vec x,t),
\end{eqnarray}
where $C$ is the counting rate [s$^{-1}$], $J_{\alpha}$ is the differential flux of the $\alpha^{th}$ kind of particle [s$^{-1}$cm$^{-2}$sr$^{-1}$E$^{-1}]$, $\varepsilon_{\alpha}$ is the detection efficiency for the $\alpha^{th}$ particle species, $t_{0}$ is the time at the start of the observation, $T$ is the total time of the observation, $d\vec\sigma$ is an element of the surface area of the detector, $d\omega = d\varphi d(\cos\theta)$ is an element of the solid angle with an azimuthal angle $\varphi$ and polar angle $\theta$, $\vec x$ is the spatial position of the detector, $\hat r$ is the unit vector in the direction $\omega$, $S$ is the total area of the detector and $\Omega$ is the domain of $\omega$. Under the special case of the detector efficiency $\varepsilon_{\alpha}$ being independent of $\omega$, $\vec\sigma$, and $t$, as well as the particle flux $J$ being independent of $\vec x$ and $t$, and an assumption that the detector only responds to a particle energy range of $E_{l}\le E\le E_{u}$, Equation~(\ref{eq:counting}) reduces to
\begin{eqnarray}
\label{eq:revcounting}
C = \int_{E_l}^{E_u}\biggr[\int_{\Omega}d\omega\int_{S}d\vec\sigma\cdot\hat r F(\omega)\biggr]dEJ_o (E),
\end{eqnarray}
where $F(\omega)$ is the angular dependence of the intensity with $F(\omega)=1$ representing isotropic fluxes. The geometric factor of the detector is the expression in square brackets in Equation~(\ref{eq:revcounting}). For the cases of circular or rectangular aperture geometries, the geometric factor has been derived from this expression (\citealt{thomas1972}). In this paper, the quantity in the bracket will be denoted as the geometric factor. However, it may be equivalently called the response function (\citealt{sullivan1971}; \citealt{thomas1972}).

In Equation~(\ref{eq:revcounting}), it is assumed that the particle trajectories are straight after entering the aperture of the telescope to the detector. Complication arises as the quantity in the bracket is often energy dependent. The intense magnetic field and energetic particles from the Sun induce physical events near the Earth such as the acceleration of particles in radiation belts, occasionally leading to damage to, or loss of, artificial satellites (\citealt{lanzerotti2001}; \citealt{baker2002}). Under such high-radiation environment, the energies of the particles are sufficiently high to allow penetration of, or scattering off, the structure before detection by the sensor. Particle trajectories from the penetration or scattering off the structure can be generated depending on the incident energies, which has been ignored in the analytic calculation. The geometric factor of the detector can be significantly affected by those penetrating or scattered particles because high-energy particles outside the Field-of-View (FOV) can be detected together with the particles within FOV.

In this research, a numerical method is suggested to calculate the geometric factor of the detectors relative to the isotropic fluxes of electrons and positive ions. The method was developed with the {\tt GEometry ANd Tracking 4 (GEANT4)} toolkit and is applied to the output from a model instrument. An isotropic flux is generated from the surface of a hypothetical sphere encompassing the detector to simulate space radiation. The initial positions and momenta of the detected particles are analyzed to identify whether they have physically interacted with the structure of the instrument prior to the final detection by the detector. It is shown in this study that the response of the detector can be as a function of the incident energies. This method could be used to conveniently analyze a more realistic geometric factor consisting of contributions from the cleanly entering particles, together with penetrating or scattered particles.

Section~\ref{sec:nummet} describes the numerical methods used to simulate the isotropic particle environment. The analytic geometric factor calculation and the method used to classify the detected particles are also explained in this section. Derived responses of the detector as compared with the analytic geometric factor are described in Section~\ref{sec:resul}, followed by Section~\ref{sec:conc} which presents the conclusions of this paper.

\section{Numerical Methods\label{sec:nummet}}

\subsection{Initial Condition\label{sec:initcon}}

It is assumed in this study that the incident particle fluxes are isotropic because such conditions are often found in space (\citealt{wilson1991}; \citealt{ncrp2006}; \citealt{durante2011}). To simulate an isotropic flux, electrons and protons are produced from the surface of a hypothetical sphere which is sufficiently large to encompass the model instrument in this investigation (\citealt{wilson2005}; \citealt{ersmark2006}; \citealt{yando2011}; \citealt{martinez2012}; \citealt{zhao2013}).
There are three inputs needed for particle generation on the surface of the hypothetical sphere; the initial values of 1) the position vector ($x_{pos}$, $y_{pos}$, $z_{pos}$), 2) the momentum vector ($p_x$, $p_y$, $p_z$), and 3) the energy ($E_{init}$). Each value of the position, momentum, and energy of a particle has been obtained randomly based on the assumptions described below. In order to achieve the uniform creation of particles over the full surface of the hypothetical sphere, the cosine value of the positional polar angle $\theta_{pos}$ ($\cos\theta_{pos}$) is uniformly given in the range $-1\le\cos\theta_{pos}\le+1$, while the positional azimuth angle $\varphi_{pos}$ is uniformly allocated in the range $0\le\varphi_{pos}\le2\pi$. The positional polar angle $\theta_{pos}$ and positional azimuth angle $\varphi_{pos}$ are later converted to the Cartesian coordinate position vector ($x_{pos}$, $y_{pos}$, $z_{pos}$). In what follows, we call the frame in which calculations are performed be called the “$S$ frame.”

The particle momentum vector ($p_x$, $p_y$, $p_z$), as expressed in the same coordinate system as the positional vector, is related to the momentum vector ($p_x'$, $p_y'$, $p_z'$) in the so-called “$S'$ frame”, in which the origin is at the position vector ($x_{pos}$, $y_{pos}$, $z_{pos}$) and the $z'$-axis lies along the line from ($0$, $0$, $0$) to ($x_{pos}$, $y_{pos}$, $z_{pos}$). The vector ($p_x'$, $p_y'$, $p_z'$) is transformed to the momentum vector ($p_x$, $p_y$, $p_z$) according to the following relation
\begin{eqnarray}
\label{eq:momentmat}
\left(\begin{array}{l}p_x\\p_y\\p_z\end{array}\right) = - \left(\begin{array}{ccc} \cos\varphi_{pos} & -\sin\varphi_{pos} & {0} \\ \sin\varphi_{pos} & \cos\varphi_{pos} & {0} \\ {0} & {0} & {1} \end{array} \right) \nonumber\\ \left(\begin{array}{ccc} \cos\theta_{pos} & {0} & \sin\theta_{pos} \\ {0} & {1} & {0} \\ -\sin\theta_{pos} & {0} & \cos\theta_{pos} \end{array} \right)
\left(\begin{array}{l}p_x'\\p_y'\\p_z'\end{array}\right).
\end{eqnarray}
%

\begin{figure}[t!]
\centering
\includegraphics[width=84mm]{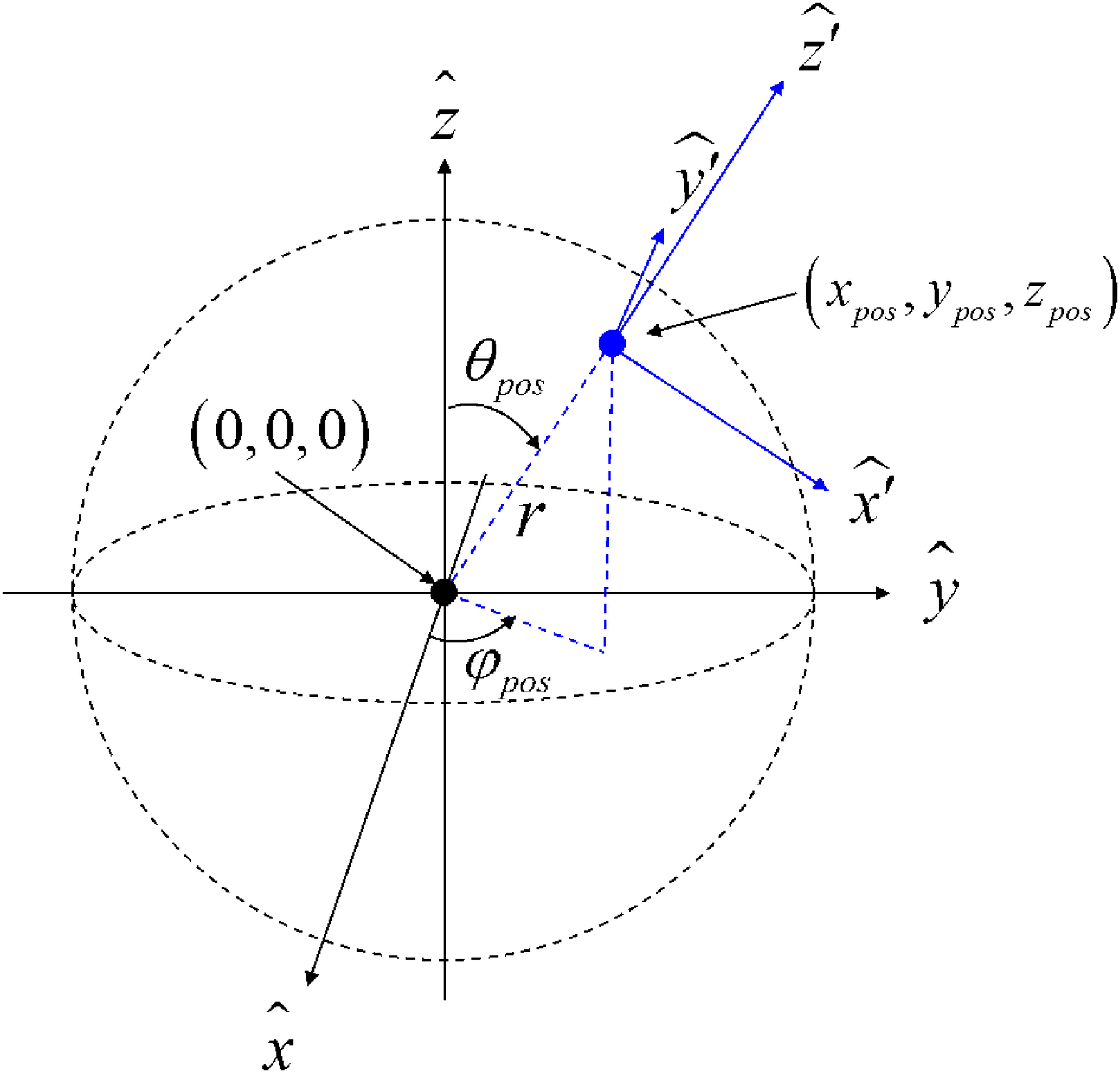}
\caption{The relation between the $S$ frame (black) and $S'$ frame (blue). The Cartesian coordinate ($x'$, $y'$, $z'$) in the $S'$ frame is the same as the spherical coordinate ($\theta$, $\varphi$, $r$) in the $S$ frame.\label{fig:fig1}}
\end{figure}

To produce isotropic fluxes within the sphere from each initial position of the particles, the half sine-squared value of the directional polar angle $\sin\theta'$ ($\sin^2\theta'/2$) is uniformly generated in the range [$0$, $1/2$], and the directional azimuth angle $\varphi'$ is uniformly drawn from the range [$0$, $2\pi$] (\citealt{zhao2013}).

\begin{figure}[t!]
\centering
\includegraphics[width=84mm]{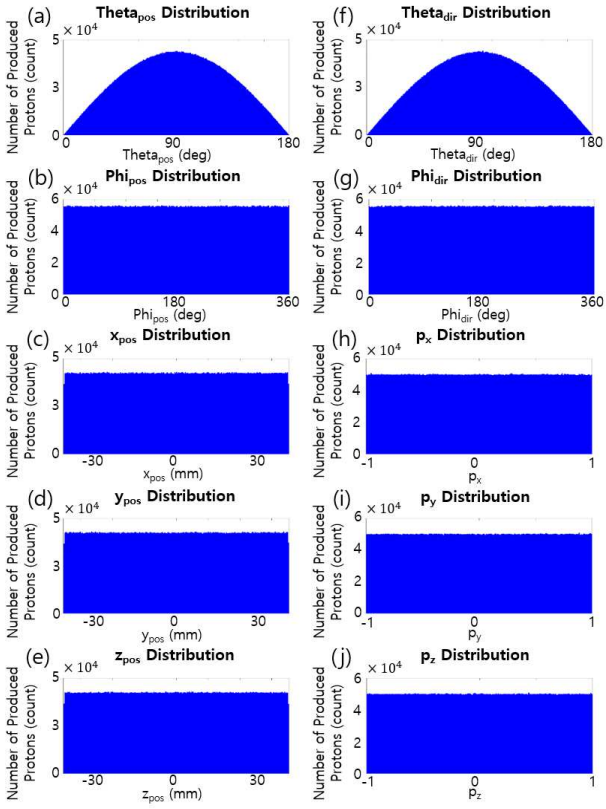}
\caption{Histograms of the initial positions and propagation directions of created protons. The distribution of generated $\theta_{pos}$ and $\varphi_{pos}$ are shown in (a) and (b). $x_{pos}$, $y_{pos}$ and $z_{pos}$ transformed from $\theta_{pos}$ and $\varphi_{pos}$, are shown in (c), (d) and (e). These histograms show that the distribution of position vectors in the Cartesian coordinates, $x_{pos}$, $y_{pos}$ and $z_{pos}$, is uniform in the range $-25\sqrt 2$~mm to $25\sqrt2$~mm. The distribution of propagation direction, $\theta_{dir}$ and $\varphi_{dir}$, transformed from $\theta'$ and $\varphi'$, is similarly illustrated in (f) and (g), with $p_x$, $p_y$ and $p_z$, shown in (h), (i) and (j). The distributions of $p_x$, $p_y$ and $p_z$, which are Cartesian components of the unit momentum vector, are uniform in the range $-1$ to $+1$.\label{fig:fig2}}
\end{figure}

\begin{figure}[t!]
\centering
\includegraphics[width=84mm]{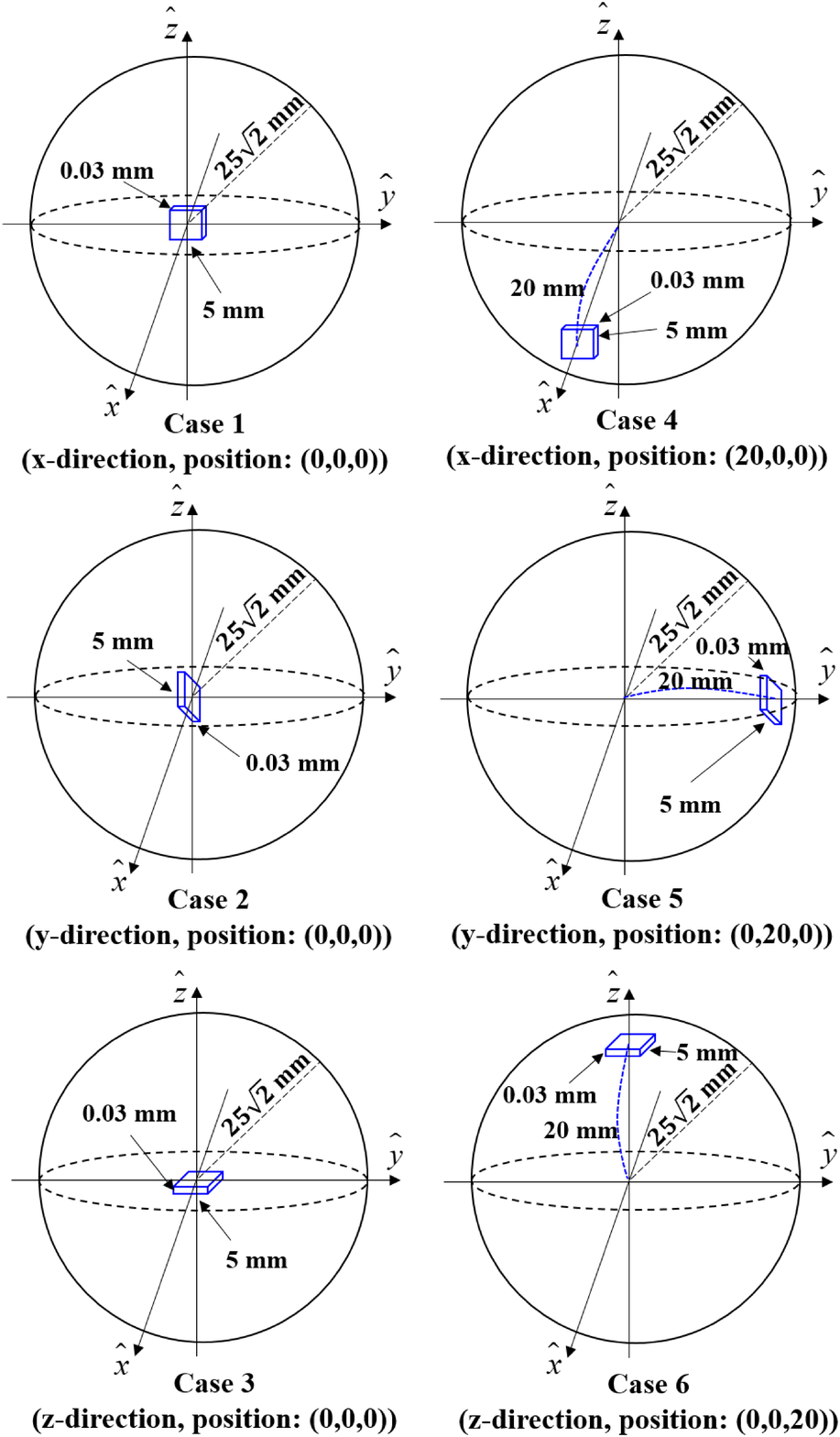}
\caption{Six simulation cases for verification of the homogeneity and isotropy of the generated proton flux within the hypothetical sphere. A single silicon detector is placed at the origin and oriented along each axis for Cases 1 to 3, while the detector is moved $20$~mm from the origin along each axis for Cases 4 to 6. From Case 1, 2 and 3, it can be verified whether the environment in the hypothetical sphere is independent of the direction, and the existence of position dependence can be checked with Cases 4, 5 and 6.\label{fig:fig3}}
\end{figure}

We assume that the energy distribution of the incident particles is found below 10~MeV for electrons and 50~MeV for protons, because this investigation is intended for near-earth space where such energy distributions are occasionally found (\citealt{piet2006}; \citealt{suparta2014}; \citealt{borovsky2017}). Extension of this method to higher energies should be straightforward. A uniform distribution of initial energies in the logarithmic scale is obtained in the energy range $10\, \mathrm{keV} \le E_{init} \le 10\, \mathrm{MeV}$ for electrons and $10\, \mathrm{keV} \le E_{init} \le 50\, \mathrm{MeV}$ for protons. The initial energy $E_{init}$ is defined as
\begin{eqnarray}
\label{eq:inite}
E_{init} = 10^k ~ \rm [keV],
\end{eqnarray}
where the random number $k$ is uniformly drawn from the range $1 \le k \le 4$ for electrons and $1 \le k \le 4.69$ for protons.

The following analysis has been undertaken to verify the isotropic and homogeneous radiation environment with the given position vector ($x_{pos}$, $y_{pos}$, $z_{pos}$) and momentum vector ($p_x$, $p_y$, $p_z$). Ten million ($1\times10^7$) protons with a fixed energy of 1~MeV are sampled from the surface of the sphere. The radius of the sphere is chosen to be $25\sqrt2$~mm to ensure efficiency in simulation time. The initial positions and momenta of all created protons are exhibited as histograms in Figure~\ref{fig:fig2}. In Figure~\ref{fig:fig2}, it is confirmed that the particles are uniformly generated from the surface of the sphere by observing that the $x_{pos}$, $y_{pos}$ and $z_{pos}$ distributions of the initial positions are uniform. The components of the vector $p_x$, $p_y$ and $p_z$ also have a uniform distribution, as seen in the right panels of the figure.

\begin{table*}[t!]
\caption{Comparison between the predicted count ($n_1$) for isotropic radiation and simulated count ($n_2$).\label{tab:tab1}}
\centering
\begin{tabular}{lcccc}
\toprule
 & & Predicted count & Simulated count & Percent difference \\
 & Direction & $(n_1 = \pi AI)$ & $(n_2)$ & $\left( \frac{|n_1-n_2|}{(n_1+n_2)/2}\times 100 \right)$ \\
 & & [counts] & [counts] & [\% difference] \\
\midrule
Case 1-(1) & X & 32\,213 $(\pm 180)^{*}$ & 32\,269 & $0.17$ \\
Case 2-(1) & Y & 32\,213 $(\pm 180)$ & 31\,992 & $0.69$ \\
Case 3-(1) & Z & 32\,213 $(\pm 180)$ & 31\,838 & $1.17$ \\
Case 1-(2) & X & 96\,638 $(\pm 311)$ & 96\,693 & $0.06$ \\
Case 2-(2) & Y & 96\,638 $(\pm 311)$ & 96\,079 & $0.58$ \\
Case 3-(2) & Z & 96\,638 $(\pm 311)$ & 96\,426 & $0.22$ \\
Case 1-(3) & X & 161\,063 $(\pm 402)$ & 160\,991 & $0.04$ \\
Case 2-(3) & Y & 161\,063 $(\pm 402)$ & 160\,803 & $0.16$ \\
Case 3-(3) & Z & 161\,063 $(\pm 402)$ & 161\,392 & $0.20$ \\
Case 4 & X & 32\,213 $(\pm 180)$ & 32\,302 & $0.27$ \\
Case 5 & Y & 32\,213 $(\pm 180)$ & 32\,495 & $0.87$ \\
Case 6 & Z & 32\,213 $(\pm 180)$ & 32\,023 & $0.59$ \\
\bottomrule
\end{tabular}
\tabnote{
  The difference between the predicted count and simulated count decreases as the number of protons is increased. The simulated counts are independent of the position of the detector, as described in Cases 4, 5 and 6 compared to Case 1-(1), Case 2-(1) and Case 3-(1), respectively.
  \\ $^{*}$Values in parentheses are Poisson uncertainties $(\sqrt{n_1})$.
}
\end{table*}

Six cases are presented in Figure~\ref{fig:fig3} which are selected to verify the anticipated homogeneity and isotropy of the simulated radiation environment. A single silicon detector is placed at the center of the sphere facing each axis for Cases 1 to 3, whereas the detector is displaced 20~mm from the center along each axis for Cases 4 to 6. For Cases 1, 2 and 3, $1\times10^7$, $3\times10^7$ and $5\times10^7$ particles are created to ascertain the cumulative error in a specific direction. These cases are called Case 1-(1), Case 1-(2), Case 1-(3), Case 2-(1), Case 2-(2), Case 2-(3), Case 3-(1), Case 3-(2) and Case 3-(3), respectively. For Cases 4, 5 and 6, $1\times10^7$ particles are generated. The predicted count ($n_1$) for isotropic radiation and simulated count ($n_2$) for each of the cases are shown in Table~\ref{tab:tab1}. In Table~\ref{tab:tab1}, $A$ denotes the active area ($A = 0.506\, [\mathrm{cm}^2]$) of the silicon detector and $I$ denotes a proton flux of $I = N/4\pi^2R^2$, where $N$ is the number of generated protons and $R = 25\sqrt2$~mm. Table~\ref{tab:tab1} shows that the percent differences are less than 1.5\% between all cases and decreases as more particles are simulated. Table~\ref{tab:tab1} also indicates that the count is independent of the position of the detector. The difference between detected counts and expected counts from theory is commensurate with Poisson statistics.

The only difference in the initial condition of the simulated particles presented in the next section is that the radius of the hypothetical sphere becomes larger to encompass the instrument. All other values remain the same in the simulation for the model instrument.

\subsection{Configuration\label{sec:confi}}

{\tt GEometry ANd Tracking 4 (GEANT4)} is a {\tt C++} toolkit for simulating physical interactions between particles and matter in the presence of an electromagnetic field (\citealt{agostinelli2003}; \citealt{ivanchenko2004}; \citealt{allison2007}). A variety of physical models are offered in this toolkit for different specific research purposes (\citealt{chauvie2004}; \citealt{valentin2012}). In this simulation, the Penelope model is used for simulating Compton scattering, Rayleigh scattering, the photoelectric effect, the Bremsstrahlung process, ionization and pair production (\citealt{salvat2003}; \citealt{sempau2003}). For the purpose of increasing accuracy in the presence of multiple scattering events of protons and electrons, two multiple scattering models \textendash\, the Wentzel-VI model and Goudsmit-Saunderson MSC model \textendash\, are adopted (\citealt{ivanchenko2010}; \citealt{fioretti2017}).

In this study, an omni-directional particle simulation is performed by {\tt GEANT4 version 9.4 patch 01}, and compilation is performed by {\tt gcc 4.5.1}, {\tt a Class Library for High Energy Physics (CLHEP) 2.1.0.1} and {\tt gnu make 3.82}. The structure of the model instrument is designed by {\tt SOLIDWORKS 2015 3D Computer Aided Design (CAD)} in the {\tt STandard for the Exchange of Product (STEP)} model data format and is converted by {\tt Fastrad 3.4.3.0} to the {\tt Geometry Description Markup Language (GDML)} format for input to {\tt GEANT4} (\citealt{kim2012}; \citealt{poole2012}; \citealt{park2014}). Five hundred million ($5\times 10^8$) electrons, and the same number of protons, are produced in this simulation, following the initial conditions described in Section~\ref{sec:initcon}.

\subsection{Geometric Factor\label{sec:geofac}}

The instrument used in this study has a detector stack inside and a collimator which has physical volume to restrict the FOV. The aperture placed on the exterior of the collimator, pointing to outer space is called the outer aperture, while the one facing the detector stack is named the inner aperture. The actual instrument used in this simulation has a more realistic geometry that is intended for space flight. A complete description of the complicated detector geometry is not provided in this paper. Such a description is unnecessary as this work mainly focuses on numerical methods and analysis of the results, given the input detector geometry that is necessary to run the simulation.

Based on the telescope geometry, the analytically calculated geometric factor is $2.2\times10^{-2}\, \mathrm{cm}^2\,\mathrm{sr}$ (\citealt{sullivan1971}; \citealt{thomas1972}). This is the ideal response $\tilde{G}$ of the detector, determined only by the FOV of the instrument with an assumption of straight particle trajectories based on analytic calculations. The geometric factor $G$, which includes the contributions of particles from outside the FOV due to particle penetration, scattering or other physical mechanisms, can be calculated via Monte Carlo simulations (\citealt{wu1988}; \citealt{jun2002}). The response of the model instrument under isotropic flux over the full $4\pi$ steradians of space is assumed to be (\citealt{yando2011}; \citealt{zhao2013}; \citealt{park2014})
\begin{equation}
\label{eq:geo}
G = \frac{n}{j_0} ~ .
\end{equation}
Here, $j_0$ is the flux of simulated particles and $n$ is the count of simulated hits on the detector. As the number of total incident particles is $N$ and the particle producing area is the surface of a hypothetical sphere of radius $r$, Equation~(\ref{eq:geo}) is converted to the following form
\begin{equation}
\label{eq:trangeo}
G = \frac{n}{N}\,4\pi^2r^2 ~ .
\end{equation}
In this research, results are expressed by the geometric factor $G$ as a function of particle initial energy $E_{init}$, which is compared with the ideal geometric factor $\tilde{G}$.

\subsection{Data Classification\label{sec:datacla}}

\begin{figure}[t!]
\centering
\includegraphics[width=70mm]{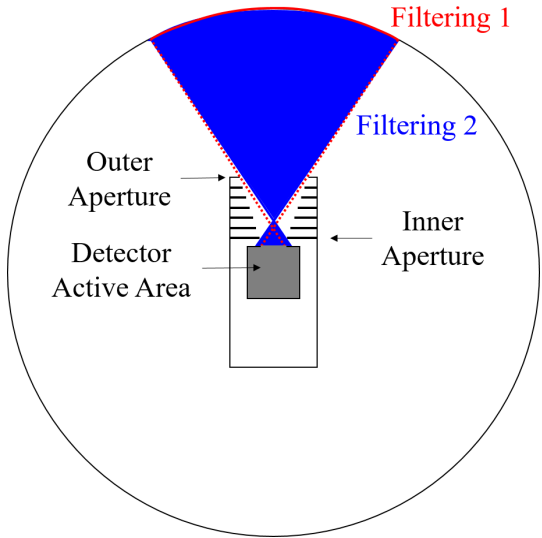}
\caption{Illustration of the filtering process for identification of detected counts in terms of initial position and momentum. Hypothetical lines (dotted in red) from the vertices of the detector stack to the vertices of the collimator outer aperture are drawn to define the FOV of the instrument. The projected area of the hypothetical shooting sphere is drawn as a red arc (designated as filtering process 1). The shaded area in blue is defined to select momentum vector that would yield direct detection in the absence of any interaction with the instrument structures (designated as filtering process 2).\label{fig:fig4}}
\end{figure}

The particles which trigger the detector stack are categorized by two steps according to the initial position ($x_{pos}$, $y_{pos}$, $z_{pos}$) and the initial momentum ($p_x$, $p_y$, $p_z$). The first step is to check whether the detected particles are generated within the assigned FOV of the telescope. For finding the surface of the hypothetical sphere which corresponds to the FOV, hypothetical lines from the vertices of detector stack to the vertexes of the collimator outer aperture are drawn and extrapolated to the surface of the hypothetical sphere. The surface area restricted to the FOV is analogized from the coordinate values of the fixed points. Let the detected particles with an initial position in the area of the FOV be called ``FOV particles'' with the rest be called ``non-FOV particle''. By this process (filtering 1), we can check whether particles penetrate the physical structure before hitting the active area of the detector. The second step (filtering 2) is to verify whether the FOV particles will directly travel to the inner aperture of the collimator, which is the last structure before the active area. This filtering process is illustrated in Figure~\ref{fig:fig4}.

\begin{figure}[t!]
\centering
\includegraphics[width=84mm]{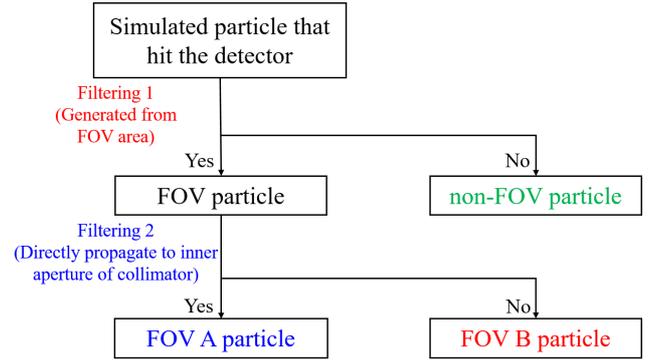}
\caption{The classification of simulated particles that hit the detector. A particle generated from the surface area restricted by the FOV of aperture is called an FOV particle, while one created from any other area is named a non-FOV particle. In the group of FOV particles, the particles directly propagating to the inner aperture of the collimator are designated as FOV A particles. The rest of the particles are classified as FOV B particles.\label{fig:fig5}}
\end{figure}

Each of the initial direction vectors combined with the initial position vectors of the FOV particles are compared with the coordinate values of the inner aperture of the collimator. If the FOV particle directly propagates to the inner aperture, we name it an ``FOV A particle'', and if not, it is an ``FOV B particle''. We can interpret whether FOV particles are scattered or not according to this process. See Figure~\ref{fig:fig5} for a schematic description of this classification process.

\section{Results and Discussions\label{sec:resul}}

\begin{figure}[t!]
\centering
\includegraphics[width=84mm]{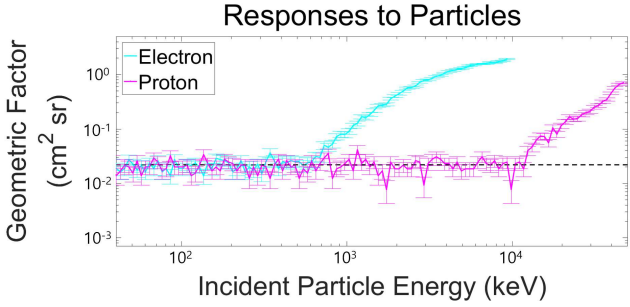}
\caption{The detector response to particles. The geometric factor of the detector stack is shown with incident particle energy. The responses to the low-energy protons and electrons below 800~keV are similar to that from the analytic geometric factor because most of the detected particles enter the telescope within the assigned FOV. On the contrary, the responses to electrons over 800~keV and protons over 10~MeV are significantly increased. This is mainly due to the penetrating particles as explained in Section~\ref{sec:resul}.\label{fig:fig6}}
\end{figure}

The overall responses to the particles are illustrated in Figure~\ref{fig:fig6}. The ideal geometric factor $\tilde{G}$ is shown as a black dashed line in the figure. The responses to electrons and protons are drastically increased in the high energy range.

\begin{figure}[t!]
\centering
\includegraphics[width=84mm]{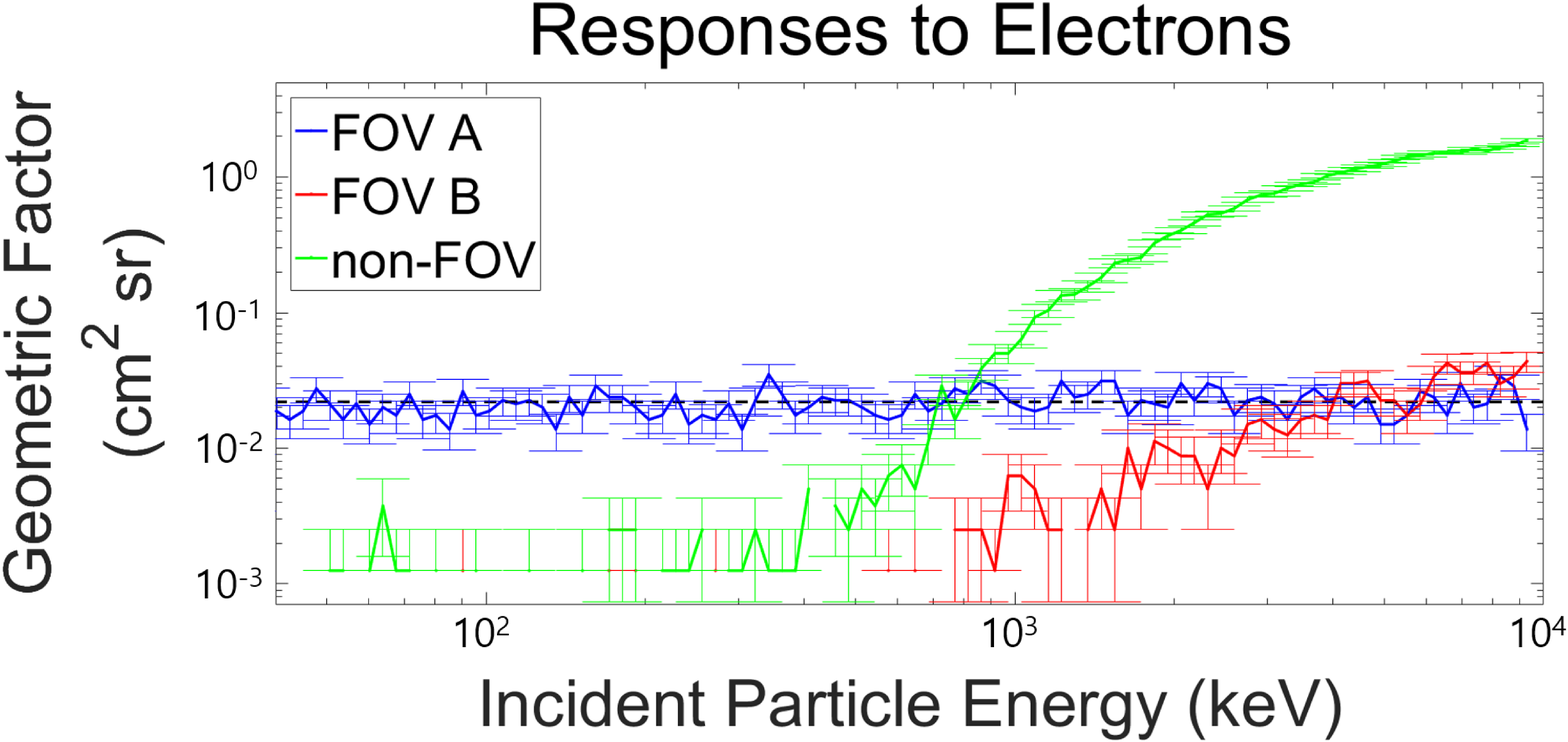}
\caption{The detector response to electrons. The geometric factor for FOV A electrons closely matches the analytic geometric factor. The responses to penetrating or scattered electrons correspond to the geometric factors for FOV B and non-FOV electrons.\label{fig:fig7}}
\end{figure}

The responses to the electrons analyzed by the FOV method are exhibited in Figure~\ref{fig:fig7}. The figure clearly demonstrates how the particles contribute to the total responses of the detector in terms of the particle energies. The responses due to particles that do not interact with the physical structure of the telescope, FOV A electrons, are similar to the analytic geometric factor. The increase of the geometric factor in the high energy range is largely due to non-FOV electrons. The contributions from FOV B electrons in the low energy range are negligible, whereas those in the high energy range are comparable to or greater than the contributions from FOV A electrons. It is interpreted that electrons classified as FOV B are largely associated with scattering processes because direct detection of such electrons is not possible without modification of the initial trajectories. The non-FOV electrons found mostly in the energy ranges above $\sim$1~MeV are detected through the process of scattering, penetration or combination thereof. There is a body of literature that reports the detection of these unwanted particles, both electrons and protons, from in-situ measurements in space (\citealt{rodger2010}; \citealt{ni2011}; \citealt{turner2012}; \citealt{asikainen2013}). The accuracy of the data interpretation critically relies on the understanding of the detector responses with respect to the incident particle energies.

\begin{figure}[t!]
\centering
\includegraphics[width=84mm]{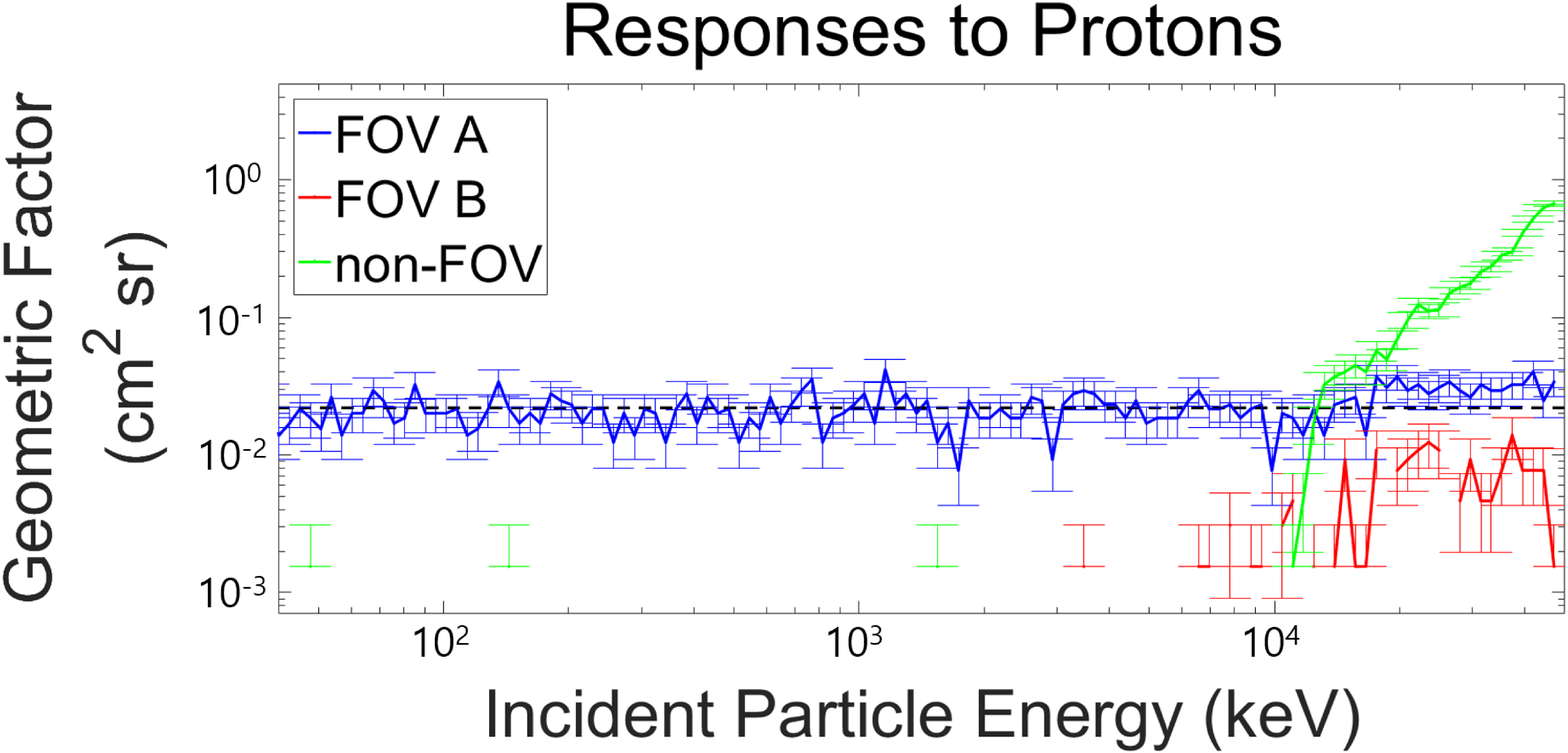}
\caption{The detector response to protons. The response to FOV A protons is similar to the analytic geometric factor. The response to non-FOV protons significantly increase in the high energy range, while the geometric factor for FOV B protons is low and can be neglect. \label{fig:fig8}}
\end{figure}

The responses to the protons, using the same FOV classification as electrons, are illustrated in Figure~\ref{fig:fig8}. The responses to FOV A protons again correspond to the ideal response. While the responses to FOV B electrons are sufficiently significant to be comparable to the geometric factors for FOV A electrons in the high energy range, the responses of FOV B protons are not because the large proton mass does not allow significant scattering off the detector structures. The response to non-FOV protons penetrating the structure of the instrument steeply increase in the energy range greater than 10~MeV.

\section{Conclusions\label{sec:conc}}

A numerical method has been introduced to calculate detector response relative to isotropic fluxes of particles. This study suggests a method to classify particles using their initial positions and momenta. Not only can the ideal geometric factor of instruments be deduced from the response to FOV A particles, but the effects from penetrating or scattered high energy particles also can be estimated from the response to FOV B and non-FOV particles.

The operating orbit of an instrument should be accounted for when assessing the significance of identified responses to high energy, non-FOV particles. For instance, the geosynchronous orbit has remarkably weak particle fluxes in high energy ranges (\citealt{nagatsuma2017}). Hence, it is expected that non-FOV particles would not make substantial contributions to the total number of observations when the model instrument is operating in this orbit. In addition, this analysis will also provide an important clue to the accurate understanding of detector responses in the presence of intense relativistic electrons in the radiation belt when the geomagnetic space is disturbed. Regardless of local experimental conditions, the calculation of the geometric factor from omni-directional particle fluxes is valuable to predict real responses across a variety of space weather events. This method is a useful tool to analyze the response from omni-directional particle simulations, and it additionally provides an easy way to calculate and understand the complicated geometric factor. We expect that the proposed method will be readily used for other instruments as well to calculate the geometric factor of detectors, and to analyze and understand the output data from in-orbit instruments.


\acknowledgments

This work was supported by the BK21 plus program through the National Research Foundation (NRF) funded by the Ministry of Education of Korea.





\end{document}